\documentclass[review]{elsarticle}

\usepackage{lineno,hyperref}
\usepackage{mathtools}
\usepackage{mathptmx}
\usepackage{float}
\usepackage{slashed}
\usepackage[usenames]{color}
\usepackage{xcolor}
\modulolinenumbers[5]

\journal{IJMP E}

\begin{document}

\begin{frontmatter}

\title{Lower bound for the neutrino magnetic moment from kick velocities induced at the birth of neutron stars}

\author{Alejandro Ayala$^{1,2}$, Santiago Bernal-Langarica$^{1}$, S. Hern\'andez-Ortiz$^{1,3}$, L. A. Hern\'andez$^{1,2,4,5}$ and D. Manreza-Paret$^{1,6}$}

\address{$^1$Instituto de Ciencias Nucleares, Universidad Nacional Aut\'onoma de M\'exico, Apartado Postal 70-543, CdMx 04510, Mexico.\\
$^2$Centre for Theoretical and Mathematical Physics, and Department of Physics, University of Cape Town, Rondebosch 7700, South Africa.\\
$^3$Institute for Nuclear Theory, University of Washington, Seattle, WA, 98195, USA.\\
$^4$Departamento de F\'isica, Universidad Aut\'onoma Metropolitana-Iztapalapa, Av. San Rafael Atlixco 186, CdMx 09340, Mexico.\\
$^5$Facultad de Ciencias de la Educaci\'on, Universidad Aut\'onoma de Tlaxcala, Tlaxcala, 90000, Mexico.\\
$^6$Facultad de F\'isica, Universidad de La Habana, San Lazaro y L, La Habana, Cuba.}

\begin{abstract}
	We show that the neutrino chirality flip, that can take place in the core of a  neutron star at birth, is an efficient process to allow neutrinos to anisotropically escape, thus providing a to induce the neutron star kick velocities. The process is not subject to the {\it no-go theorem} since although the flip from left- to right-handed neutrinos happens at equilibrium, the reverse process does not take place given that right-handed neutrinos do not interact with matter and therefore detailed balance is lost. For simplicity, we model the neutron star core as being made of strange quark matter. We find that the process is efficient when the neutrino magnetic moment is not smaller than $4.7 \times 10^{-15}\mu_B$, where $\mu_B$ is the Bohr magneton. When this lower bound is combined with the most stringent upper bound, that uses the luminosity data obtained from the analysis of SN 1987A, our results set a range for the neutrino magnetic moment given by $4.7 \times 10^{-15} \leq \mu_\nu/\mu_B \leq (0.1 - 0.4)\times 10^{-11}$. The obtained kick velocities for natal conditions are consistent with the observed ones and span the correct range of radii for typical magnetic field intensities.
\end{abstract}


\end{frontmatter}


At the endpoint of stellar evolution lies a class of astrophysical compact objects, such as White Dwarfs (WD), Neutron Stars (NS) and Black Holes (BH) whose physical properties still remain intriguing at large. WD were first discovered more than one hundred years ago, whereas NS and BH appeared in the astronomical panorama only approximately fifty years ago. For this class of objects, the Newtonian theory of gravity fails, and efforts to confirm their existence as well as to understand their properties,  occupied much of the twentieth century's research in astrophysics. In fact, the main theories in physics formulated over the last century, such as general relativity, statistical physics, nuclear and particle physics,  hydrodynamics as well as astrophysics, have converged in the study of this kind of objects and at the same time, these theories have received feedback from the study of compact objects.

The internal composition of WD and NS is of particular interest since their extremely high densities cannot be reproduced in terrestrial laboratories. Thus, they represent physical realizations of  systems to study the phases of matter that are as yet not well understood. In particular, the extreme conditions of density, temperature and intensity of magnetic fields present in NS have been the source of the great interest that such objects arise. These conditions have led research on this field to propose the existence of strange and exotic phases of matter within the internal layers of NS. The current efforts to map the phase diagram of nuclear matter in the high baryon density, low temperature domain, have found in NS a source of information that heavy-ion experiments have little or no chance to explore.

An important aspect in the study of NS is the evolution of their properties with time. This aspect is connected with neutrino emission which drives the NS cooling properties. In fact the continuous leaking of neutrinos is responsible for the NS drastic change of temperature from dozens to almost zero MeV in a time lapse of about one hundred years from the NS formation. Moreover, if this emission of neutrinos is  anisotropic, it can also be responsible for the so called pulsar kicks~\cite{Hobbs:2005yx,Lai:2000pk,Chugai}.

Several scenarios have been proposed to explain the unusual pulsar velocity distribution, including the hydrodynamic scenario, the electromagnetic rocket effect and the anisotropic neutrino emission. The main mechanisms are reviewed in \cite{Wang_Lai}. 
One of these is a natal kick, which can be due to processes occurring during the formation of the proto-NS whereby a burst of neutrinos in a preferred direction can provide the NS with momentum in the direction opposite to the emitted flux of neutrinos. Another possibility is a post-natal kick whereby this flux of neutrinos mainly happens over the course of the rest of the NS time evolution.
An argument in favor of the natal kick scenario is provided in Refs.~\cite{Janka2017, 2019ApJ...876..151S} where calculations show the existence of a peak in the electron neutrino luminosity at an early time, right after the core bounce stage for the NS formation that lasts for about 12 seconds. During this stage, the NS temperature is about 50 MeV. Another possible scenario to explain NS kicks is the formation of quark matter during the collapse~\cite{Sumiyoshi_2017, Keranen:2004vj}.

In a previous work, we computed the NS kick velocities resorting to a scenario whereby the neutrino anisotropic emission is caused by the presence of a magnetic field in the NS interior~\cite{PhysRevD.97.103008}. However, since the neutrino mean free path is small compared to the core radius, scattering can cause that the neutrino emission does not happen along a preferred direction, thereby diluting the effect. In this work, we complement that study by introducing the idea of a neutrino chirality flip produced by the existence of a neutrino magnetic moment, which, as pointed out in Ref \cite{Fujikawa}, can exist for massive neutrinos. When neutrinos flip their chirality becoming right-handed from the original left-handed state, their interactions with matter become suppressed. Thus, even if the flip happens in equilibrium, right-handed neutrinos cannot be converted back into left-handed ones and detailed balance is lost, which makes the no-go theorem discussed in Ref.~\cite{kusenko} to become non-applicable. Moreover, in the presence of a magnetic field, right-handed neutrino resonance convertion back into left-handed ones is suppressed in the NS core where the matter effects dominate~\cite{1999PhRvD..59k1901A, Ayala:1999xn}. In this manner, the neutrinos can escape from the NS, provided that the path length for the chirality flip process happens within one mean free path. We show that this requirement places a lower bound for the magnitude of the neutrino magnetic moment which is larger than the Standard Model (SM) value for this magnetic moment. We show that this result, combined with the most stringent upper bound for the neutrino magnetic moment~\cite{1999PhRvD..59k1901A, Ayala:1999xn} leaves a window of values to make the process plausible.

The scenario we discuss is set to happen during the transition from the so called stage III into stage IV, according to the main stages of NS evolution described in Ref.~\cite{Lattimer536}. This implies that the kick velocity would be driven by a neutrino burst that starts around 15 seconds after the core collapse, at the beginning  of  the  NS  lifetime, and ends approximately 35 seconds later.
As discussed in Refs.~\cite{refId0}, temperatures below 1 MeV won't produce observable velocities and, according to Ref.~\cite{Lattimer536}, these temperatures are reached around the first minute. Therefore, the main contribution to the kick velocities comes from the neutrino emission at large temperatures (between 50 and 30 MeV) and chemical potentials, at the beginning of the NS evolution.

We thus study the chirality flip process within the core of the NS assuming these conditions which are consistent with those assumed to place the upper bound on the neutrino magnetic moment from the chirality flip in supernovae. Following the procedure described in Refs.~\cite{1999PhRvD..59k1901A,Ayala:1999xn}, to compute the chirality flip rate we consider that when neutrinos are being produced, the plasma is in thermal equilibrium at temperature $T$ and with an electron chemical potential $\mu_e$ such that $T,\mu_e \gg m_e$, where $m_e$ is the electron mass. The production rate of a right-handed neutrino, with energy $p_0$ and momentum $\vec{p}$ from left-handed neutrinos is given by
\begin{equation}
\label{neuchir1}
    \Gamma (p_0) = \frac{\tilde{f}(p_0)}{2p_0} \text{Tr}\left[\slashed{P}R\text{Im}\Sigma\right],
\end{equation}
where $\tilde{f}(p_0)$ is the Fermi-Dirac distribution for right-handed neutrinos, taken for simplicity as massless, such that $P_\mu = (p_0,\vec{p})$, $\|\vec{p}\| = p$. The operator $R= \frac{1}{2}(1+\gamma_5)$ projects onto right-handed fermion components. The neutrino self-energy $\Sigma$ is given by
\begin{equation}
\label{neuchir2}
\Sigma(P)=T\sum_n \int \frac{d^3 k}{(2\pi)^3} V ^\rho (K) \, S_F(\slashed{P} - \slashed{K})\, L \, V ^\lambda (K) \, D_{\rho\lambda} (K),
\end{equation}
where $K_\alpha = (k_0,\vec{k})$, $\|\vec{k}\|=k$, $V^\mu$ is the neutrino-photon vertex function, $S_F$ is the neutrino propagator and $D_{\rho\lambda}$ is the photon propagator. For the neutrino-photon vertex we use the magnetic dipole interaction, given by $V _\mu(K) = \mu_\nu \, \sigma_{\alpha\mu}K^\alpha$, where $\mu_\nu$ is the neutrino magnetic moment and $\sigma_{\alpha\mu} = \frac{i}{2} \left[\gamma_\alpha,\gamma_\mu\right]$. The photon propagator is split into longitudinal and transverse components
\begin{equation}
\label{neuchir3}
D_{\rho\lambda} (K) = \Delta_L (K) P^L _{\rho \lambda} + \Delta _T (K) P^T _{\rho \lambda}.
\end{equation}
The corresponding sums over Matsubara frequencies are
\begin{equation}
M_{L,T}=T\sum_n \Delta_{L,T} (i\omega_n)\tilde{\Delta}_F (i(\omega-\omega_n)).
\end{equation}
In order to evaluate the sums, we introduce the photon and neutrino spectral densities. Therefore, the imaginary part of $M_{L,T}$ can be written as
\begin{align}
\mathrm{Im}\left[M_{L,T}\right]=\pi\left( e^{\beta (p_0 - \mu)} + 1 \right)& \int_{-\infty} ^\infty \frac{dk_0}{2\pi} \int_{-\infty} ^\infty \frac{dp'_0}{2\pi} f(k_0)\tilde{f}(p'_0 - \mu) \nonumber \\
&\times \delta(p_0 - k_0 - p'_0) \rho_{L,T} (k_0,k) \rho_F (p'_0),
\label{neuchir6}
\end{align}
where $f(k_0)$ is the Bose-Einstein distribution, $\rho_{L,T}$ and $\rho_{F}$ are the spectral densities for the photon and neutrino propagators, respectively, given by
\begin{align}
\label{neuchir7}
\rho_F(p'_0)&=2\pi \epsilon(p'_0) \delta({p'_0} ^2 - E_p ^2),\nonumber\\
\rho_L(k_0,k)&=\frac{x}{1-x^2}\frac{2\pi m_\gamma^2\theta(k^2-k_0^2)}{\left[k^2+2m_\gamma^2\left(1-\frac{x}{2}\ln |(1+x)/(1-x)|\right)\right]^2+\left[\pi m_\gamma^2x\right]^2} ,
\nonumber\\
\rho_T(k_0,k)&=\frac{\pi m_\gamma^2x(1-x^2)\theta(k^2-k_0^2)}{\left[k^2(1-x^2)+m_\gamma^2\left(x^2+\frac{x}{2}(1-x^2)\ln \left|(1+x)/(1-x)\right|\right)\right]^2+\left[(\pi/2)m_\gamma^2x(1-x^2)\right]^2},
\end{align}

where we have defined $x =k_0/k$ ,  $\epsilon(p'_0) $ is the sign function
and the photon thermal mass $m_\gamma$ is given by
\begin{eqnarray}
  m_\gamma^2=\frac{e^2}{2\pi^2}\left(\mu_e^2 + \frac{\pi^2 T^2}{3}\right).
\label{photonnmass}
\end{eqnarray}
Using Eq.~\ref{neuchir6}, one finds that the expression for $\Gamma$ involves
the factor
\begin{align}
\label{neuchir9}
 &\pi\left( e^{\beta (p_0 - \mu)} + 1 \right)\int \frac{d^3 k}{(2\pi)^3}\int_{-\infty} ^\infty \frac{dk_0}{2\pi} f(k_0) \left(\sum_{i=L,T}C_i \rho_i (k_0)\right) \nonumber \\
&\times \frac{1}{2E_p}\left[ \tilde{f}(E_p - \mu)\delta(E_p - (p_0-k_0))   - \left(1-\tilde{f}(E_p + \mu)\right)\delta(E_p - (p_0-k_0)) \right],
\end{align}
where $C_L$ and $C_T$ are functions that come from the trace calculation and the contraction with the longitudinal and transverse polarization tensors respectively. These are given by
\begin{eqnarray}
    C_L &=& -k^2 \left(1-x^2\right)^2 \left(2p_0 - k_0\right)^2 ,\\
    C_T &=& k^2 \left(1-x^2\right)^2 \left[\left(2p_0 - k_0\right)^2 - k^2\right].
\end{eqnarray}
The term proportional to $\tilde{f}(E_p - \mu)$ in Eq.~\eqref{neuchir9}, corresponds to the case where the left-handed neutrino is in the initial state, while the term proportional to $\left(1-\tilde{f}(E_p + \mu)\right)$ represents the case where this neutrino is in the final state. Since we seek to describe the production of  right-handed neutrinos from initially left-handed neutrinos, we must keep only the contribution from the first term.

After performing the angular integrals, we can finally express $\Gamma$ as
\begin{eqnarray}
    \label{neuchir10}
    \Gamma(p_0)&=&  \frac{\mu_\nu ^2}{32\pi^2 p_0 ^2} \int_0 ^\infty dk \, k^3 \int_{-k} ^{k} dk_0\ \theta(2p_0 +k_0 - k) \nonumber \\
    & & \times \: [1+f(k_0)]\, \tilde{f}(p_0 + k_0 -\mu) \, (2p_0 + k_0)^2 \left(1-\frac{k_0^2}{k^2}\right)^2 \nonumber \\
    & & \times \: \left[ \rho_L (k_0) + \left( 1 - \frac{k^2}{(2p_0 + k_0)^2}\right) \rho_T (k_0) \right].
\end{eqnarray}

Armed with the expression for the reaction rate as a function of $p_0$, we can compute the total reaction rate as the integral of $\Gamma(p_0)$ over the available phase space, namely,  \begin{equation}
    \Gamma= V\int\frac{d^3p}{(2\pi)^3}\Gamma(p_0)
    =\frac{V}{2\pi^2}\int_0^{p_0^{max}}dp_0\ p_0^2\ \Gamma(p_0),
    \label{totalrate}
\end{equation}
with $V$ the volume where the chirality flip process takes place and we integrate up to the maximum energy allowed for the neutrino in the beta decay process. Since beta decay is mediated by an intermediate $W$ whose mass is much larger than the typical energy scales during the proto-NS birth, we take $p_o^{max}\simeq 1.2$ MeV, which is the maximum value for a massless neutrino energy for beta decay in vacuum. The typical time required for the chirality flip process to take place is given as the inverse of $\Gamma$, namely
\begin{eqnarray}
   \tau = 1/\Gamma.
    \label{tau}
\end{eqnarray}
For an efficient anisotropic emission, this time should be smaller than the time required for a neutrino to travel one mean free path $\lambda$ at the speed of light $c$ namely
\begin{eqnarray}
   c\tau\leq \lambda.
   \label{condition}
\end{eqnarray}
This condition can be used to place a lower bound for the neutrino magnetic moment above which the chirality flip process can efficiently produce neutrinos to  escape from the NS and thus serve as a mechanism to induce the kick. For this purpose, let us consider the changing temperature for conditions of a NS made out of SQM~\cite{PhysRevD.58.083001} during its evolution from the initial temperature $T_i=50$ MeV to the final temperature $T_f=30$ MeV. The neutrino chemical potential and the electron chemical potential --entering the expression for the photon thermal mass-- evolve with temperature. Notice that these conditions are also used for the calculation of the upper bound of the neutrino magnetic moment in Refs.~\cite{1999PhRvD..59k1901A,Ayala:1999xn} from luminosity data obtained from the analyses of SN 1987A. To account for the neutrino emission during the extended temperature interval we use the {\it average} interaction rate defined as the integral of Eq.~(\ref{totalrate}) with respect to the temperature, divided by the temperature interval. We take $\lambda\sim 1$ m, The reaction volume is taken as a cylinder with longitudinal axis equal to the mean free path and base with radius equal to the magnetic field coherence length $l_m$, namely,
\begin{equation}
    \label{volcylinder}
    V = \pi \, l_m ^2 \, \lambda .
\end{equation}
For the strong field limit ($B \sim 5\times 10^{16}$ G), $l_m$ is approximately given by $l_m \approx 2000$
fm~\cite{LATTIMER2007109}.

\begin{figure}[t!]
    \centering
    \includegraphics[width=200pt]{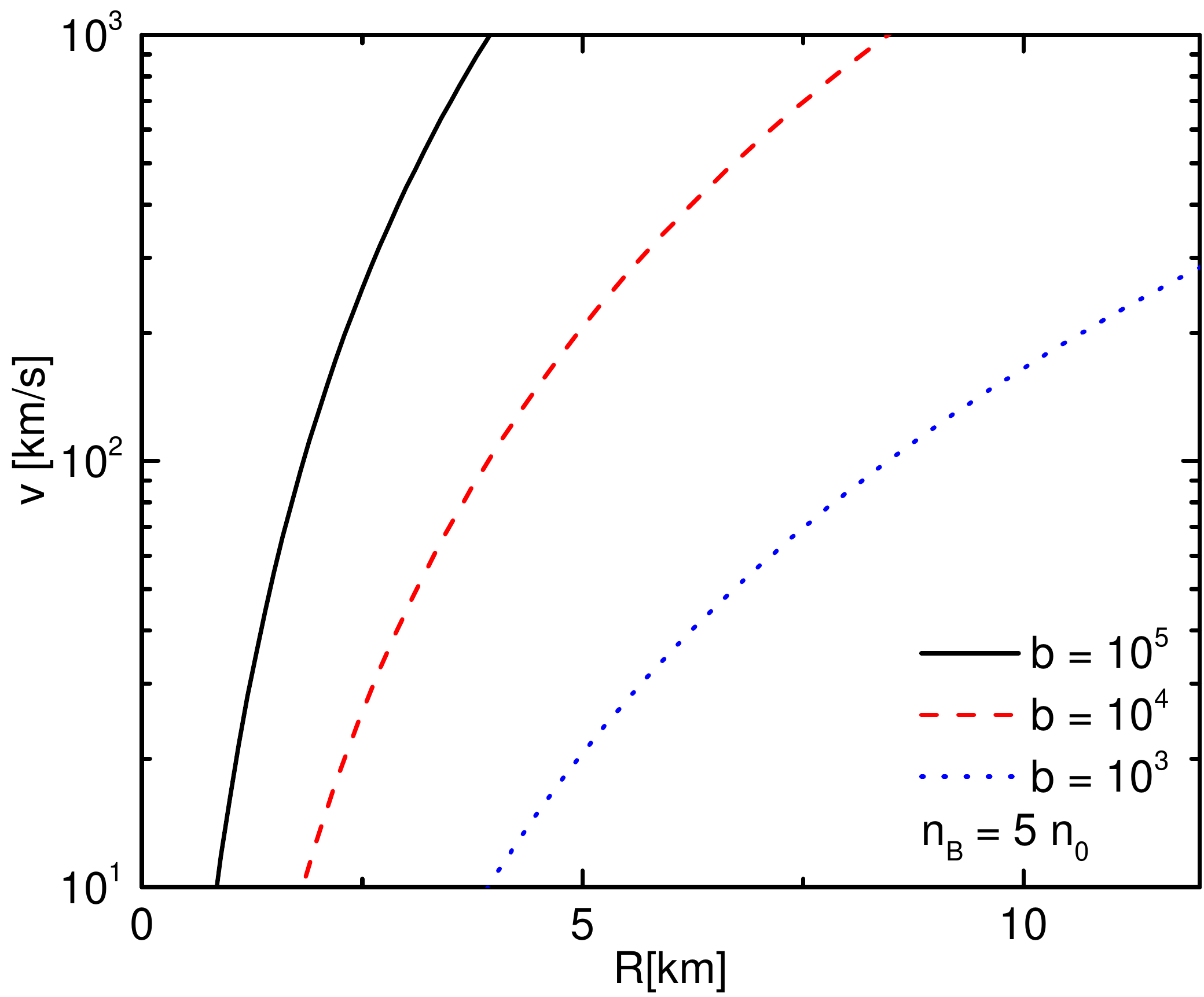}
     \caption{Kick velocities as a function of the NS radius for a NS with 1.4 solar masses, for different values of the magnetic field $B=10^{16} - 10^{18}$ G, corresponding to values of the ratio of the magnetic field to the critical magnetic field $b\sim 10^3-10^5$. The obtained kick velocities are within the observed ones.}
    \label{fig1}
\end{figure}

Putting all this information together, we obtain the lower bound for the neutrino magnetic moment
\begin{equation}
    \label{HFbound}
     4.7 \times 10^{-15}\mu_B\leq \mu_\nu,
\end{equation}
where $\mu_B$ is the Bohr magneton. Combining this result with the upper bound for the neutrino magnetic moment obtained from a similar analysis applied to the core collapse during a supernova explosion, we obtain the range for the neutrino magnetic moment given by
\begin{eqnarray}
   4.7 \times 10^{-15} \leq \mu_\nu/\mu_B \leq (0.1 - 0.4)\times 10^{-11}.
   \label{range}
\end{eqnarray}
This result can be contrasted with the bound obtained in a minimum extension of the SM allowed by the non-vanishing neutrino mass~\cite{Marciano:1977wx,Lee:1977tib}. Using the neutrino parameters deduced from solar, atmospheric and reactor neutrino experiments, this bound turns out to be~\cite{Balantekin:2006sw}
\begin{eqnarray}
 4\times 10^{-20} \leq\mu_\nu /\mu_B,
   \label{SMlimit}
\end{eqnarray}
which is smaller than the lower bound obtained in Eq.~(\ref{range}).

When the magnetic moment is above the lower bound of Eq.~(\ref{SMlimit}) and the neutrinos flip their chirality before one mean free path, all the neutrinos escape and the produced kick velocity for NS can be computed from \cite{refId0,Sagert_2007}
\begin{equation}\label{vel_kick}
dv=\frac{\chi }{M_{NS}}\frac{4}{3}\pi R^3\epsilon dt,
\end{equation}
where $M_{NS}$ and $R$ are the NS core's mass and radius, respectively. $\epsilon$ is the neutrino emissivity and $\chi$ is the electron spin polarization, which, after all the neutrinos flip their chirality,  become functions only of temperature, density and magnetic field intensity.

We point out that since NS are compact and massive, the emission of energy form these objects in any from can receive corrections from general relativistic (GR) effects~\cite{Lattimer_GR, Page_Webber, Potekhin}. For instance, the  correction to the neutrino luminosity is proportional to the metric factor $e^{2\Phi}$ which, in the case of a typical neutron star, is around 0.5. Therefore, due to gravitational effects, the neutrino luminosity should be about one half of the estimated one. However, the  asymmetry in the emission is not affected and the energy released in neutrinos is around $10^{53}$ erg. In addition, since the neutrino mean free path is much smaller than the NS radius, there are no GR corrections to the estimate of the interaction volume and hence, for our purposes, these corrections can be neglected. For a discussion on other GR effects see Ref.~\cite{GR_effects}.

When the emissivity changes with temperature, the cooling equation can be used, namely,
\begin{equation}\label{vel_kick3}
-\epsilon=\frac{dU}{dt}=\frac{dU}{dT}\frac{dT}{dt}=C_v\frac{dT}{dt},
\end{equation}
where $U$ is the internal energy density and  $C_v$ is the heat capacity. Therefore, the kick velocity is given by
\begin{equation}\label{velchi1}
v=-\frac{1 }{M_{NS}}\frac{4}{3}\pi R^3\int_{T_i}^{T_f}\chi\, C_vdT.
\end{equation}
This velocity can be written in the following form
\begin{equation}
v=-803.925 \ \frac{\text{km}}{\text{s}}\left( \frac{1.4M_\odot}{M_{NS}}\right)\left( \frac{R}{10 \ \text{km}}\right)^3 \left( \frac{I}{\text{MeV} \ \text{fm}^{-3}}\right),
\label{velchi2}
\end{equation}
where
\begin{equation}\label{vel2}
I=\int_{T_i}^{T_f}\chi\, C_vdT.
\end{equation}

The integral  $I$ is a function  of $C_v$ and $\chi$. These two quantities depend on the strength of the magnetic field, the chemical potential and temperature. These quantities were derived and studied in~\cite{PhysRevD.97.103008}.
The heat capacity is given by $C_{v}=\sum_f C_{vf}$ where
\begin{equation}\label{Cv}
C_{vf}=\frac{d_fm^2}{4\pi^2 T^2}b_f\int_{0}^\infty dp_3\sum_{l=0}^{\infty}(2-\delta_{l0})\frac{(E_{lf}-\mu_f)^2}{[1+\cosh{\frac{E_{lf}-\mu_f}{T}}]},
\end{equation}
and $E_{lf}$ is given by
\begin{equation}\label{Disp-Rel}
E_{lf}=\sqrt{p_3^2+2|e_f|Bl+m_f^2},
\end{equation}
with $e_f$ and $m_f$ being the fermion charge and mass, respectively, $l$ the Landau level, $p_3$ the momentum along the magnetic field direction $\textbf{B}$, and $b_f=B/B_f^c$, with $B_f^c=m_f^2/e_f$ the critical magnetic field. We use the shorthand notation  $b_e\equiv b=B/B_e^c$. The electron spin polarization is given by
\begin{equation}\label{chi2}
\chi=\left\lbrace 1+ \frac{2
\sum\limits_{\nu=1}^{\infty}\int_0^\infty dx_3 \frac{1}{e^{(\frac{m_e}{T}\sqrt{x_3^2+1+2\nu b}-x_e)}+1}}{\int_0^\infty dx_3 \frac{1}{e^{(\frac{m_e}{T}\sqrt{x_3^2+1}-x_e)}+1}} \right\rbrace ^{-1},
\end{equation}
where $x_3=p_3/m_f$ is the dimensionless momentum.

For simplicity, we consider the core of the NS as a plasma made out of magnetized strange quark matter (SQM), namely, a gas composed of quarks $u$, $d$ and $s$ and electrons $e$ in the presence of a magnetic field. SQM may exist in the core of some heavy neutron stars~\cite{Annala_2020}, and we will take this type of matter as the framework for our calculations. 
Several scenarios to describe the formation of SQM inside a NS have been proposed. In one kind of scenarios~\cite{Pagliara, Witten}, the whole star makes a transition to a strange quark matter. However other possibilities have also been studied, including the existence  ~\cite{Alford, Schramm, Dexheimer}. For instance, in Ref.~\cite{Alford_Reddy}, the two simplest possibilities to describe the transition from ordinary nuclear matter to strange quark matter as the pressure increases are discussed: (1) a single sharp interface between nuclear and quark matter and (2) a mixed phase region. For our purposes, we consider the latter kind of scenarios having in mind a hybrid situation in which SQM exists only in the core of the NS.

We also impose the conditions that are believed to exist in the core of neutron stars  which are determined by $\beta$ decay equilibrium among quark species, charge neutrality, baryon number conservation and an electron plus neutrino to baryon ratio $Y_L = 0.4$ \cite{PhysRevD.52.661, Reddy:1997yr}. All together, these conditions are referred to as the stellar equilibrium conditions.

Assuming that the neutrino magnetic moment is within the range given by Eq.~(\ref{range}), the anisotropic neutrino emission becomes an efficient mechanism for the NS kick. Figure~1 shows the range of kick velocities obtained for natal core conditions and for a typical 1.4 solar mass NS and different values of the magnetic field with a fixed central baryon density $n_B=5n_0$, where $n_0$ corresponds to ordinary nuclear density. Notice that the range of velocities is in agreement with the observed ones, provided the NS radius lies within the given range. Notice also that larger magnetic field intensities correspond to larger kick velocities and that the upper limit for observed velocities is obtained for small star radii.

In conclusion, we have shown that neutrino chirality flip during the birth of NS, whose core has been modeled as consisting mainly of SQM, is an efficient mechanism to allow the produced neutrinos to  escape from the NS core, provided the neutrino magnetic moment is not smaller than $4.7 \times 10^{-15}\mu_B$. When this lower bound is combined with the most stringent upper bound~\cite{1999PhRvD..59k1901A,Ayala:1999xn}, the results of this work set a range for the neutrino magnetic moment given by $4.7 \times 10^{-15} \leq \mu_\nu/\mu_B \leq (0.1 - 0.4)\times 10^{-11}$. Notice that this lower bound  is in agreement with the one found in Ref.~\cite{Akhmedov}, where neutrinos acquire a transition magnetic moment when they resonate between different flavors allowing them to escape from the NS.

The obtained kick velocities for natal NS conditions are consistent with the observed ones and span the correct range of radii for typical magnetic field intensities. Nevertheless, they are computed under the simplifying assumptions that the chirality flip happens before the neutrinos travel less than one mean free path and that the core of the NS is made out primarily of SQM. However, in principle, the probability of this flip is distributed along the neutrino trajectory such that a fraction of neutrinos can travel more than one mean free path without having flipped their spins. This in turn translates into a fraction of neutrinos not escaping from the core of the NS in a preferred direction which impacts the neutrino emissivity and thus the kick velocity, making it smaller. Matter in the core of the NS, although possibly containing SQM, needs also to be treated incorporating more realistic equations of state (EoS). In this sense, the results of this work have to be regarded as providing upper limits for the kick velocities. A more complete study, where the chirality flip distribution along the neutrino trajectory is considered and more sophisticated EoS are used, is currently under way and will be reported elsewhere.

When neutrinos escape from the recently born NS by the mechanism advocated in this work, conditions inside the core, over the period when its gravitational binding energy is radiated away, can change. These changes would need to be incorporated into the modern transport calculations~\cite{Janka2017}. These studies promise to open a window to connect different observational aspects of NS properties, some of which are planned to be addressed in a future work.

\section*{Acknowledgements}
This work was supported by Consejo Nacional de Ciencia y Tecnolog\'ia grant numbers A1-S-7655 and A1-S-16215 and by UNAM-DGAPA-PAPIIT grant number IG100219. L. A. H. acknowledges support from a PAPIIT- DGAPA-UNAM fellowship. S. H. O. acknowledges the support from the U.S. DOE under Grant No. DE-FG02-00ER41132 and the Simons Foundation under the Multifarious Minds Program Grant No. 557037.

\end{document}